\documentclass{svjour2}                    % onecolumn
\smartqed  % flush right qed marks, e.g. at end of proof
\usepackage{graphicx}
\usepackage{mathptmx}      % use Times fonts if available on your TeX system
\def\be{\begin{eqnarray}}
\def\ee{\end{eqnarray}}

\journalname{Foundations of Physics}
\begin{document}

\title{Measurement-based quantum computation and undecidable logic}


\author{Maarten Van den Nest         \and
        Hans J. Briegel %etc.
}


\institute{M. Van den Nest \at
              Institut f\"ur Quantenoptik und Quanteninformation der \"Osterreichischen Akademie der Wissenschaften, Innsbruck, Austria\\
              Tel.: +43-512507-4751\\
              Fax: +43-512507-9815\\
              \email{maarten.van-den-nest@uibk.ac.at}
          \and
           H. J. Briegel \at
              Institut f\"ur Quantenoptik und Quanteninformation der \"Osterreichischen Akademie der Wissenschaften, Innsbruck, Austria\\
          Institut f{\"u}r Theoretische Physik, Universit{\"a}t Innsbruck,
Technikerstra{\ss}e 25, A-6020 Innsbruck, Austria}
                      %  \\

\date{Received: date / Accepted: date}

\maketitle

\begin{abstract}
We establish a connection between measurement-based quantum computation and  the field of mathematical
logic. We show that the computational power of an important class of
quantum states called \emph{graph states}, representing
resources for measurement-based quantum computation, is reflected in
the expressive power of (classical) formal logic languages defined
on the underlying mathematical graphs. In particular, we
show that for all graph state resources which can yield a computational
speed-up with respect to classical computation, the underlying
graphs---describing the quantum correlations of the states---are
associated with \emph{undecidable} logic theories. Here
undecidability is to be interpreted in a sense similar to G\"odel's
incompleteness results, meaning that there exist
propositions, expressible in the above classical formal logic, which
cannot be proven or disproven.
\keywords{Quantum information theory\and Quantum computation \and Logic}
\end{abstract}

\section{Introduction}
\label{intro}

Quantum computers are devices that
use quantum mechanics for enhanced ways of information
processing  \cite{Fe85}. Indeed, it is known that problems such as integer factoring can be performed significantly faster on a quantum computer than on any known classical device  \cite{Sh97}. Despite these exciting perspectives, the questions:
\begin{center}
\emph{ ``What are the essential resources that give quantum computers their computational power''},

and

\emph{``Are quantum computers fundamentally more powerful than classical devices?''}
\end{center}
remain to date largely unanswered.  
%\cite{Jo03} \cite{Vi03}, \cite{Ma05}, \cite{Sh05}, \cite{Jo06}, \cite{Yo06}, \cite{Gr06}, \cite{Va06}, %\cite{Va06b}.

The existence of several models for quantum computation, each based on different concepts, indicates that there may not be a straightforward answer to these difficult questions. The new paradigm of \emph{measurement-based}, or {\em one-way} quantum computation  \cite{Ra01}, \cite{Ra02}  has lead to novel perspectives in these respects. The introduction of this model established that certain many-qubit quantum states, such as the \emph{2D cluster states}   \cite{Br01}, exhibit the remarkable property that universal quantum computation can be achieved by simply individually measuring the qubits of the system in a specific order and basis, and by classical processing of the measurement results.  The initial state of the system then serves as the resource for the entire computation which is (in part) consumed in the process.  This is in sharp contrast to the quantum circuit model, where computations are realized via unitary evolution.   Within the  measurement-based paradigm for quantum computation, fundamental questions regarding the speed-up of quantum with respect to classical computation can be formulated and investigated in an alternative, and in several cases much more concise, way. In particular, the introductory questions of this paper can be restated as
\begin{center}
\emph{``Which resource states for measurement-based quantum computation (MQC) yield a computational speed-up over classical computers?''}
\end{center}
This question will be adressed in the present article.

As entanglement can only decrease in a one-way computation,
the enhanced computational power of such a quantum computer
(beyond a classical Turing machine) must originate in the
entanglement structure of its resource state. Owing to this
insight, a series of papers have recently been devoted to investigating which types of entanglement needs to be present in any resource state which achieves the desired enhanced computational power \cite{Jo03} \cite{Vi03} \cite{Ma05} \cite{Sh05} \cite{Jo06} \cite{Yo06} \cite{Gr06} \cite{Va06} \cite{Va06b}. In this paper  we establish a new necessary condition for resource states to yield a computational speed-up with respect to classical computers. The crucial point of this result is that the present criterion is entirely different in nature with respect to previously established requirements---in particular, it will \emph{not} be stated in terms of entanglement. In the following we will focus on resource states belonging to the rich class of \emph{graph states}, which are generalizations of the 2D cluster states and which play an important role in several applications in quantum information theory (e.g., one-way quantum computation, quantum error-correction, multipartite entanglement theory, communication schemes; see \cite{He06} for a review). A graph state on $n$ qubits is defined by means of a mathematical graph on $n$ vertices, which completely encodes the correlations in the system. Our main result will be a connection between the possibility of obtaining a computational speed-up w.r.t. classical computation by performing MQC on   graph state resources, and
certain properties of the associated graphs in relation with mathematical logic theory---more
particularly,  the \emph{decidability} of logic theories.

As is well known since G\"odel's incompleteness theorems
 \cite{Goe31}, every formal system that is sufficiently
interesting (or rich) contains statements which can neither
be proved to be true nor to be false within the axiomatic
framework of the system. G\"odel's incompleteness theorem
not only applies to formal systems relating to natural
numbers (cf. Peano arithmetic) but also to graphs. Indeed,
many interesting
graph properties can be expressed within a (classical) formal language, denoted by ${\cal L}$ (the exact definition of this language is stated below).
Examples  of such properties are planarity or
2-colorability of graphs.  In this paper  we will show that
the computational power of graph states---as resources for
measurement-based quantum computation---is reflected in the
expressive power of the formal language ${\cal L}$ defined
on the underlying graphs, which encode the set of quantum correlations in the system. In particular, the following
result will be obtained.

\

{\bf Theorem.}  \emph{Graph state resources for measurement-based quantum computation can only
yield a computational speed-up over classical computers if the formal language ${\cal L}$ defined on the
underlying graphs is undecidable.}

\

Here undecidability  is to be interpreted in a sense
similar to G\"odel, meaning that there exist propositions,
expressible in the logic ${\cal L}$, which cannot be proven
or disproven.

The Theorem  provides a necessary condition to assess
the computational power of graph state resources by
considering the underlying graphs, which, by the very
definition of graph states, render a classical
encoding of the quantum correlations in the states. The concept of
undecidability is to be regarded as a notion of complexity
of the graphs---and hence of the correlations in the
system---stated independently of any quantitative (entanglement) measure.

This paper aims at connecting two quite different fields of
research, namely quantum computation and mathematical logic.
Therefore, in the following we will give a brief review of
the basic concepts of measurement-based (one-way) quantum
computation, as well as logic theories on graphs. This will
allow us to reformulate the main theorem in a
precise manner. The proof of the Theorem is obtained
by combining our previous results regarding entanglement
width and MQC on graph states \cite{Va06b}, and a recent
graph theoretic result \cite{Cou04}. We will conclude the
paper with an interpretation of these results.

\section{Graph states and measurement-based quantum computation}

We will be concerned with a
particular class of multi-party quantum states, called
\emph{graph states} \cite{He06}, which are generalizations of the 2D
cluster states. A graph state $|G\rangle$ on $m$ qubits is
the joint fixed point (i.e., an eigenvector with eigenvalue 1) of $m$ commuting correlation operators
\be\label{K} K_a:= \sigma_x^{(a)}\bigotimes_{b\in N(a)}
\sigma_z^{(b)},\ee where $\sigma_x$ and $\sigma_z$ are
Pauli matrices, and the upper indices denote on which qubit
system these operators act. Moreover, $N(a)$ denotes the
set of neighbors of qubit $a$ in the graph $G$. Thus, a system in a graph state has $\langle K_a\rangle =1$ for every $a$. For example, a 2D cluster state is obtained if the
underlying graph is a $k\times k$ square lattice
$C_{k\times k}$ (thus $m=k^2$).

The family of 2D cluster states is known to be a \emph{universal} resource for measurement based quantum computation, in that any unitary operation can efficiently be simulated by performing measurements on a 2D cluster state of appropriate dimensions \cite{Ra01}. As graph states are generally highly entangled, and as they can
efficiently be prepared by applying a suitable poly-sized
quantum circuit to a product input state, they form natural
candidates to serve as resources for MQC. We envisage a
situation where an (infinitely large) family of graph
states \be \Psi=\{|G_1\rangle, |G_2\rangle, \dots\}\ee with
growing system size  is considered, and where local
measurements can be performed on arbitrary members of
$\Psi$, thus allowing to implement quantum computations of arbitrary length. For example, $\Psi$ could be the family of 2D cluster states, $\Psi = \{|C_{k\times k}\rangle: k=1, 2, \dots\}$.

In this setting we are interested in resources $\Psi$ for MQC which yield a computational speed-up over classical devices. Given a family of states
$\Psi$, we will say that
efficient classical simulation of MQC on $\Psi$ is possible, if for every state $|G_i\rangle \in
\Psi$ it is possible to simulate every LOCC protocol (short for \emph{local operations and classical communication}) on a classical computer with overhead poly$(m_i)$, where $m_i$
denotes the number of qubits on which the state
$|G_i\rangle$ is defined \cite{Va06b}. Evidently, resources which allow a computational speed-up over classical computers do \emph{not} allow efficient classical simulation of MQC.

In the following we will focus  on the graphs associated to
families of graph states, and the formal
languages defined on them. Note that by definition
(\ref{K}) the graph $G$ is an encoding of the correlations
present in the corresponding graph state. Therefore, any
property of these graphs reflects a property of the
corresponding states, and, more particularly, the
correlations in these states.

\section{Graphs and logic}
Next we define some basic
notions of logic theory which are necessary to state our
main results concisely below. We refer  to Refs. \cite{Cou}
\cite{Oum}  for an extensive treatment. We also emphasize that we
will favor clarity of the exposition over mathematical
rigor.

In fundamental aspects of graph theory one is interested in
formal approaches to formulate graph properties such as
2-colorability, connectedness, planarity, etc. This
formalization is obtained by defining a logical calculus in
which such graph properties can be expressed. Roughly
speaking, a logic on a graph $G$ corresponds to a set of
rules which determine the basic constituents with which
statements regarding $G$ can be constructed. Such
formalization in terms of logic allows, in principle,
artificial devices to mechanically prove or disprove, for a
given graph or set of graphs, properties expressible in
this logic.

The simplest logic is \emph{first-order logic}, which is obtained by
allowing formulas containing the following elementary components:
\begin{itemize}
\item Quantifications $\exists x$ and $\forall x$ over vertices $x$ of the graph (i.e.,
``There exists a vertex $x$ such that [...]'', or ``For all
vertices $x$ it holds that [...]'') and connectives $\wedge$,
$\vee$, and $\neg$, i.e., the logical ``AND'', ``OR'' and
``NOT''. \item Moreover, it is allowed to express whether two
vertices are adjacent in the graph or not. This is formally achieved by introducing the symbol
${\tt edge}$, which is defined by \be {\tt edge}(a, b)= \left\{
\begin{array}{cl}\mbox{True}& \mbox{if } \{a, b\}\mbox{ is an edge} \\
\mbox{False}& \mbox{otherwise},\end{array} \right. \ee for
every pair of vertices $a, b$ of the graph.
\end{itemize}
A simple example of a first-order logic formula on a graph $G$
is ``There exist vertices $x$, $y$, and $z$ such that $x$ is
connected to $y$ and $y$ is connected to $z$'' or, more
formally, \be \exists x\ \exists y\ \exists z\ {\tt edge}(x, y)
\wedge {\tt edge}(y, z).\ee 

It turns out that first-order logic is often not rich enough to
express interesting graph properties. One therefore extends
first-order logic by allowing more elementary symbols. In
particular, one may supplement first-order logic with the following
elements:
\begin{itemize}
\item Next to variables $x, y, z, ...$ denoting vertices of
the graph, one also allows \emph{set variables} $X, Y, Z, ...$
(indicated by capital letters),  which denote subsets of
vertices. \item Furthermore, one adds quantifications $\forall
X$, $\exists X$ over such sets.\item Finally, one introduces
elementary formulas of the form $x\in X$, which allow one to express
that a vertex $x$ belongs to a certain subset $X$.
\end{itemize}
The logical calculus which is thus obtained is called
\emph{monadic second-order logic}, or \emph{MS logic} in short.
MS logic is strictly more expressive than first-order logic,
i.e., there are problems which can be expressed with MS logic
which cannot be expressed using only first-order logic. An
example of an MS formula on a graph is \be\exists X\ \exists Y\
\{\forall z\ (z\in X \vee z\in
Y)\ \wedge\qquad\qquad\qquad\qquad\qquad \nonumber \\
\forall z\ \forall z'\ \neg{\tt edge}(z, z')\vee\neg(z,
z'\in X\vee z, z'\in Y) \}.\nonumber\ee This formula
expresses that the graph can properly be colored with 2
colors (i.e., the vertices can partitioned in two classes
such that no two adjacent vertices are in the same class).
Many interesting graph properties can be expressed in MS
logic, among which there are several NP-hard problems (such
as e.g. 3-colorability), indicating that MS logic on graphs
has a considerable expressive power.

A slight extension of MS logic is  obtained by including
atomic formulas of the form ${\tt Even}(X)$, indicating
that the set $X$ has even cardinality. In this way one
obtains MS logic with the additional possibility to count
modulo two, denoted by \emph{C$_2$MS logic} \cite{Cou04},
which will be our topic of interest, i.e., it corresponds
to the logic ${\cal L}$ as denoted above. C$_2$MS logic is
an interesting extension of MS logic which is moreover physically interesting in that it e.g. allows
to express whether two graph states are local unitary
equivalent \cite{foot2}.

\section{Decidability of logic theories}
Next we
introduce the fundamental notion of \emph{decidability} of
logic theories. We again refer to \cite{Cou} \cite{Oum}  for
details. Let ${\cal G}=\{G_1, G_2, \dots\}$ be a (finite or
infinite) family of graphs, and let ${\cal L}$  denote
C$_2$MS logic. The ${\cal L}$-theory of ${\cal G}$ is defined to be the collection of all formulas $\varphi$, expressed in the logic ${\cal L}$, which are satisfied (or ``True'') for \emph{all} graphs in the family ${\cal G}$. The set ${\cal G}$ is said to have a \emph{decidable } ${\cal L}$-theory if for every formula
$\varphi$ expressed in the logic ${\cal L}$, it is possible
to decide (in finite time) whether or not $\varphi$ belongs to the ${\cal L}$-theory of ${\cal G}$. The set ${\cal G}$ is said to have an
\emph{undecidable } ${\cal L}$-theory if it does not have a
decidable ${\cal L}$-theory.

For example, the formula $\varphi$ could correspond to graph
planarity, 2-colorability, etc. Then the question is asked
whether all graphs in a given family of graphs ${\cal G}$
are planar, 2-colorable, etc. If every  possible such
question, that is, every formula expressible in the
language ${\cal L}$, can be answered in finite time, then
this family is said to have a  decidable  ${\cal
L}$-theory. The decidability or undecidability of a logic
theory ${\cal L}$ on a set ${\cal S}$ is a reflection of
both the expressive power of the logic ${\cal L}$--- ``How
many properties can be expressed in the logic ${\cal
L}$?''--- and the complexity (regarded in a colloquial sense)
of the family ${\cal G}$---``How rich is the structure of
the graphs in ${\cal G}$?''.

Before giving examples of (un)decidable MS theories,  is
important to make the following two remarks. First,
decidability of a logic theory is not concerned with the
\emph{efficiency} with which problems can be solved---one
only asks whether it is \emph{in principle} possible to
verify whether a given formula $\varphi$ is true, where one
does not care about e.g. the computational complexity of a
possible verification algorithm. Second, note that
\emph{any} first-order or MS theory is decidable on
\emph{finite} families of graphs ${\cal G}=\{G_1, G_2,
\dots, G_N\}$. This is simply because, in this finite
regime, any formula can be verified by an exhaustive
enumeration of cases. Thus, decidability is only relevant
when infinite structures are considered.

Let us now give some important examples.  First, let ${\cal
G}_{\mbox{\scriptsize bin}}$ be the set of all binary tree
graphs, which are regarded as so-called incidence
structures. A milestone result was obtained by Rabin, who
proved that ${\cal G}_{\mbox{\scriptsize bin}}$ has a
decidable MS theory \cite{Ra69}. This result has many
important implications in graph theory and computer
science. Further, let ${\cal G}_{\mbox{\scriptsize{2D}}}$
be the set of all 2D ($k\times k$) lattice graphs. Then
${\cal G}_{\mbox{\scriptsize{2D}}}$ has  an
\emph{undecidable} C$_2$MS theory \cite{foot3}. As final
examples, let ${\cal G}_{\mbox{\scriptsize{tri}}}$ and
${\cal G}_{\mbox{\scriptsize{hex}}}$ be the sets of all
triangular lattice graphs and hexagonal lattice graphs,
respectively, regarded as adjacency structures. Then also
${\cal G}_{\mbox{\scriptsize{tri}}}$ and ${\cal
G}_{\mbox{\scriptsize{hex}}}$ have \emph{undecidable}
 C$_2$MS theories \cite{foot4}.

\section{Main results}
Keeping in mind that the graph
states corresponding to the 2D rectangular, hexagonal and
triangular lattices have been shown to be universal
resources for MQC \cite{Va06}, the above examples already
suggest a connection between the computational power of a
family of graph states as a resource for MQC, and the
C$_2$MS logic defined on the underlying graphs. This
connection will now be fully established, as we are now in
a position to precisely state and prove the main result of
this paper.  Let ${\cal G}=\{G_1,
G_2, \dots \}$ be an (infinitely large) family of graphs
and let $\Psi({\cal G})$ be the associated family of graph
states. The main Theorem can then precisely be formulated
as follows.

\

{\bf Theorem.} \emph{If a family of graphs ${\cal G}$ has a
decidable C$_{2}$MS logic theory then  MQC performed on the graph
state resource $\Psi({\cal G})$ can classically be simulated
efficiently.}

\

Thus, this results states that any family of graphs with a decidable C$_{2}$MS logic theory \emph{cannot} give rise to a graph state resource for MQC which yields a computational speed-up as compared to classical computers.

The proof of the Theorem  is in fact quickly  obtained by invoking
previous results of the present authors and a highly nontrivial
result from graph theory. The proof has two main
ingredients (i) and (ii): 

\begin{itemize}
\item[(i)] Courcelle and Oum  \cite{Cou04} proved that every class of
graphs ${\cal G}$ which exhibits a divergence with respect to a
graph invariant called \emph{rank-width} (we refer to Ref.
\cite{Oum} for definitions) must have an undecidable C$_{2}$MS
theory.
\item[(ii)] In previous work \cite{Va06b}, the present and other authors proved that
every family of resource states $\Psi({\cal G})$ where ${\cal G}$ has a bounded rank-width, allows
an efficient simulation of MQC. 
\end{itemize}

Combining (i) and (ii) then yields the proof
of the Theorem.

\begin{figure}%[ht]
\hspace{2cm}{\includegraphics[width=0.60\textwidth]{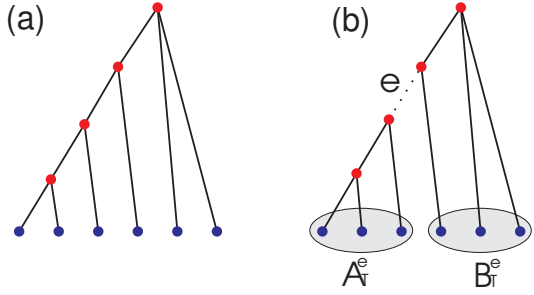}}
\caption[]{\label{subcubic} (a) Example  of a subcubic tree
$T$ with six leaves (indicated in blue). (b) Tree
$T\setminus e$ obtained by removing edge $e$ and induced
bipartition $(A_{T}^e, B_{T}^e)$.}
\end{figure}

Next we elaborate on the above proof strategy. Let us first introduce the notion ``rank-width''. The rank-width rwd($G$) \cite{Oum} of a graph $G$ is a parameter
which measures how well a graph can be approximated by means of
certain ``tree-like'' structures. Graphs with small rank-width include,
e.g., a one-dimensional chain with open or closed boundary
conditions, or a 2-dimensional ``stripe'', which is a $d \times n$
square lattice where $n$ may be arbitrarily large, but where $d$ is
held fixed. Graphs of large rank-width include e.g. 2-dimensional
$n\times n$ lattices (for growing $n$), or lattices of higher
dimensions. For completeness, we give here the definition of the
rank-width, which is quite technical [the reader who is not
interested in these mathematical details may skip to the next
paragraph]. Let $G$ be a graph with vertex set $V=\{1, \dots, n\}$ and edge set $E$.
Let $\Gamma$ be the $n\times n$ adjacency matrix of $G$, i.e, one
has $\Gamma_{ab}=1$ if $\{a, b\}\in E$ and $\Gamma_{ab}=0$
otherwise. For every bipartition $(A, B)$ of the vertex set $V$,
define $\Gamma( A, B)$ to be the $|A|\times |B|$ submatrix of
$\Gamma$ defined by \be \Gamma( A, B) := (\Gamma_{ab})_{a\in A, b\in
B}.\ee Let $T$ be a \emph{subcubic} tree, which is a tree\footnote{A
tree is a graph without cycles.} such that every vertex has exactly
1 or 3 incident edges (see Fig \ref{subcubic}(a)). The vertices which are incident with exactly
one edge are called the \emph{leaves} of the tree. For a given fixed graph
$G$, we will be interested in the collection of all possible subcubic trees $T$ with exactly $n$ leaves
$V:=\{1, \dots, n\}$, which are identified with the $n$ vertices of
$G$. Letting $e=\{i, j\}$ be an arbitrary edge of such a tree $T$, we denote by
$T\setminus e$ the graph obtained by deleting the edge $e$ from $T$.
The graph $T\setminus e$ then consists of exactly two connected
components, which naturally induce a
bipartition $(A_{T}^e, B_{T}^e)$ of the set $V$ (see Fig. \ref{subcubic}(b)). The rank-width of
the graph $G$ is now defined by the following optimization problem: \be \mbox{rwd}(G) = \min_T\
\max_{e\in T}\ \mbox{ rank}_2 \Gamma( A_{T}^e, B_{T}^e) . \ee Here the
minimization is taken over all subcubic trees $T$ with $n$ leaves,
which are identified with the $n$ vertices in the graph. Moreover,
$\mbox{ rank}_2 \Gamma( A, B) $ denotes the rank of the matrix
$\Gamma( A, B)$ when arithmetic is performed modulo 2.

One notices that the construction involving the subcubic trees is designed to single out a specific class  of bipartitions $(A_{T}^e, B_{T}^e)$ of the vertex set of $G$, over which the min-max optimization problem is performed. For a given tree $T$ one considers the maximum, over all edges
$e$ in $T$, of the quantity $\mbox{ rank}_2\Gamma(A_{T}^e, B_{T}^e)$; then the
minimum, over all subcubic trees $T$, of such maxima is computed.
As an example, it can be shown that the rank-width of a 1D chain is
equal to 1 (independent of the length of the chain), whereas an
$n\times n$ square lattice has rank-with of $O(n)$ and thus
increases with the size of the lattice.

While the rank-width is indeed a rather involved mathematical
concept, it turns out to be crucial for the current investigation.
In particular, the following highly nontrivial result by Courcelle
and Oum is of particular interest \cite{Cou04}:

\

\emph{Every class of graphs ${\cal G}$ with an unbounded rank-width
must have an undecidable C$_{2}$MS theory.}

\

This result connects the notion of rank-width of a class of graphs
with the decidability of the logic theory of this class. We will use
this result to prove the Theorem by juxtaposing it to a result
obtained by the present and other authors, which relates the
rank-width of a (family of) graph(s) with the computational power of the associated
(family of) graph state(s). For, it was proved in \cite{Va06b} that:

\

{\it Consider a family of graphs ${\cal G}$ with a bounded rank-width. Then the graph state resource $\Psi({\cal G})$ allows an efficient
classical simulation of MQC}.

\

%This result relates the concept of rank-width of a graph with the computational power of the associated graph state resource for MQC.
The combination of this result with the one obtained by Courcelle
and Oum, then immediately yields the proof of the Theorem.

Finally, we conclude this paragraph by remarking that the Theorem
represents a sufficient condition for a resource $\Psi({\cal G})$ to
be simulatable, but not a necessary one. For, there exist graph
state resources with an unbounded rank-width---and therefore an
undecidable C$_2$MS logic theory---for which MQC is nevertheless
simulatable; examples of such resources are given by the so-called
``toric code states'' \cite{Br06}, or graph states with logarithmically growing rank-width \cite{Va06b}.

\section{Discussion}
In the Theorem the desired
connection between measurement based quantum computation on graph states and mathematical
logic theories on the underlying graphs is fully obtained. It is the
authors' opinion that the present results should be
regarded as conceptual results, aimed at establishing a
connection between seemingly remote areas of research,
rather than yielding direct practical applications. We are
aware that assessing whether a family of graphs has a
decidable C$_2$MS theory is a formidable task, and that
logic theory itself is a dynamic area of research with difficult outstanding problems \cite{Foot}. Therefore the present results are not likely to e.g. directly
provide  new examples of states on which MQC can be
simulated efficiently. Nevertheless, we believe that our
findings present a new, and possibly deep, perspective
towards understanding the central issue of what the computational power of quantum computers with respect to classical devices is. The present connection to logic theory
offers an entirely new view on ``how complex'' states need to
be in order for them to possibly provide computational speed-ups,
next to more standard considerations regarding
entanglement. While the Theorem in fact follows from
previously obtained results regarding MQC and entanglement
width of graph states  \cite{Va06} \cite{Va06b},  the resulting logic criteria
are entirely different in nature.

Finally, one might be inclined to relate (C$_2$MS) logic
formulas defined on graphs to the content of the quantum computations (i.e.,
measurement patterns) implemented on the corresponding
graph states. As far as the authors are aware, there does not seem to be a direct
relation between the classical logic defined on graphs and
the quantum measurements---that might be associated to a quantum
logic---on the corresponding graph states. Thusfar it seems that the classical logic
theories, and the issue of their (un)decidability, are related
to assessing the complexity of the graphs, and hence of the
(correlations in the) graph states, with no direct
correspondence to quantum algorithms. However, it would be
very interesting to investigate this issue in more detail,
and we leave this as an open problem.

\begin{acknowledgements}
We thank Kris Luttmer and Sang-Il Oum for interesting discussions. This work was supported by the FWF and the European Union (QICS, OLAQUI, SCALA).
\end{acknowledgements}


\begin{thebibliography}{99}
\bibitem{foot3}
This follows from the fact that the 2D cluster graphs have
unbounded rank width \cite{Oum}, and that all families of
graphs with unbounded rank width have undecidable C$_2$MS
theories \cite{Cou04}.
\bibitem{foot4}
As ${\cal S}_{\mbox{\scriptsize{tri}}}$ and
${\cal S}_{\mbox{\scriptsize{hex}}}$ correspond to
universal resources \cite{Va06}, these families have
unbounded rank width. Using \cite{Cou04} then again yields
the undecidability of the corresponding C$_2$MS theories.
\bibitem{Foot}
A conjecture treated in Ref. \cite{Cou04} is particularly relevant in the present context.
\end{thebibliography}



\end{document}